\documentclass[twocolumn,showpacs,preprintnumbers,amsmath,amssymb,superscriptaddress,nofootinbib]{revtex4}
\usepackage{graphicx}

\newcommand{\eps}{\varepsilon}

\newcommand{\upU}{{\rotatebox[origin=c]{180}{$U$}}}

\newcommand{\psic}{{\Psi}}
\newcommand{\xc}{{\tilde{c}}}

\begin{document}
\title{Separability of Hamilton--Jacobi and Klein--Gordon Equations 
in General Kerr--NUT--AdS Spacetimes}

\author{Valeri P. Frolov}

\email{frolov@phys.ualberta.ca}

\affiliation{Theoretical Physics Institute, University of Alberta, Edmonton,
Alberta, Canada T6G 2G7}

\author{Pavel Krtou\v{s}}

\email{Pavel.Krtous@mff.cuni.cz}

\affiliation{Institute of Theoretical Physics, Charles University, V
Hole\v{s}ovi\v{c}k\'ach 2, Prague, Czech Republic}

\author{David Kubiz\v n\'ak}

\email{kubiznak@phys.ualberta.ca}

\affiliation{Theoretical Physics Institute, University of Alberta, Edmonton,
Alberta, Canada T6G 2G7}

\affiliation{Institute of Theoretical Physics, Charles University, V
Hole\v{s}ovi\v{c}k\'ach 2, Prague, Czech Republic}

\date{November 22, 2006}  

\begin{abstract}
We demonstrate the separability of the Hamilton--Jacobi and scalar field equations in general
higher dimensional Kerr--NUT--AdS spacetimes. No restriction on the parameters characterizing 
these metrics is imposed.
\end{abstract}

\preprint{Alberta-Thy-15-06}

\maketitle

\section{Introduction}

The study of the separability of the Hamilton--Jacobi and the
corresponding scalar field equations in a curved spacetime has a long
history. Robertson \cite{Rob} and Eisenhart \cite{Eis} discussed
general conditions for such a separability in spaces which admit a
complete set of mutually orthogonal families of hypersurfaces. An
important class of 4-dimensional separable spacetimes, including
several type D metrics, was found by Carter \cite{Carter:68}.
Carter also proved the separability of the Hamilton--Jacobi and the 
scalar field equation in the Kerr metric \cite{Carter:68a}. It was
demonstrated in \cite{Penr:70} that this separability follows from the existence of a
Killing tensor. This result was generalized later, namely, it was
shown that Killing and Killing--Yano tensors play an important role in
the separability theory (see, e.g.,
\cite{HuSo:73,Wood:75,BeFr:79,BeFr:80,PenrFl:73}).

In the present paper we prove the separability of the Hamilton--Jacobi
and scalar field equations in the general ($D\ge 4$) Kerr--NUT--AdS spacetimes
\cite{CLP}. These solutions were
obtained as a generalization of the metrics for the rotating higher
dimensional black holes with a cosmological constant
\cite{Chong:2004hw,Gibbons:2004prl,Gibbons:2004uw}, which, in their turn, are
generalizations of the Myers--Perry solution \cite{MP}. There are several
publications devoted to the separation of variables for this class of
metrics. However, all the results obtained up to now assume either a restriction on
the number of dimensions \cite{FrSta,FrStb} or special properties of
the parameters which characterize the solution 
\cite{Vasu:04a,Vasu:04b,Vasu:05,Chong:2004hw,ChenLuPope:2006,Davis:2006hy,Ort:03}.
The separation of variables which we prove in this paper is valid in the
general Kerr--NUT--AdS spacetime in any number of dimensions and
without any restriction on the parameters of the metric. 
We also discuss the relation of the separation constants with the
conserved quantities connected with the Killing--Yano and Killing
tensors recently discovered for this class of the metrics \cite{FK,KF,PKVK,QuadrConst}.

\section{Kerr--NUT--AdS metrics and their properties}

Our starting point is the general higher dimensional
Kerr--NUT--AdS metric obtained in \cite{CLP}. 
The metric can be written using ${D}$ coordinates ${x^a}$
which naturally splits into two groups.
Radial and latitude coordinates are denoted as 
${x_\mu}$ and labelled by the Greek indices 
${\mu,\nu=1,\dots,n}$, ${n=[D/2]}$, i.e., ${x_\mu=x^\mu}$.
Time and azimuthal coordinates ${\psi_k=x^{n+1+k}}$ are indexed 
by the Latin indices from the middle of the alphabet, ${k,l=0,\dots,m}$, 
${m=D-n-1}$. We use the Einstein summation convention only for the
indices ${a,b,\dots}$ running over all coordinates. For the 
convenience we also introduce ${\eps=D-2n}$.
In these coordinates the metric and its inverse read\footnote{%
The form \eqref{metric} of the metric is actually an analytical continuation 
related to the physical metric by a simple Wick rotation, see \cite{CLP}.} 
\begin{equation}\label{metric}
ds^2\!\!=\!\!\sum_{\mu=1}^n\Bigl[\frac{dx_{\mu}^2}{Q_{\mu}}
  +Q_{\mu}\!\Bigl(\sum_{k=0}^{n-1} A_{\mu}^{(k)}d\psi_k\!\Bigr)^{\!2}\Bigr]
  -\frac{\eps c}{A^{(n)}}\Bigl(\sum_{k=0}^n A^{(k)}d\psi_k\!\Bigr)^{\!2} \!,
\end{equation}
and
\begin{equation}\label{invmetric}
\begin{split}
(\partial_s)^2&=\sum_{\mu=1}^n\Bigl[ Q_\mu(\partial_\mu)^2
  +\frac1{Q_\mu U_\mu^2}\Bigl(\sum_{k=0}^m(-x_\mu^2)^{n\!-\!1\!-\!k}\partial_k\Bigr)^{\!2}\Bigr]\\
  &\mspace{220mu}-\frac{\eps}{cA^{(n)}}(\partial_n)^2 \;.
\end{split}\raisetag{4ex}
\end{equation}\\[-2ex]
Here,  
\begin{gather}
Q_{\mu}=\frac{X_{\mu}}{U_{\mu}},\quad 
U_{\mu}=\prod_{\substack{\nu=1\\\nu\ne\mu}}^{n}(x_{\nu}^2-x_{\mu}^2),\quad c=\prod_{k=1}^m a_k^2, \nonumber\\ 
X_{\mu}=(-1)^{1-\eps}\frac{1+\lambda x_{\mu}^2}{x_{\mu}^{2\eps}}\prod_{k=1}^{m}(a_k^2-x_{\mu}^2)
+2M_{\mu}(-x_{\mu})^{1-\eps},\nonumber\\
A_{\mu}^{(k)}=\!\!\!\!\!\sum_{\substack{\nu_1<\dots<\nu_k\\\nu_i\ne\mu}}\!\!\!\!\!x^2_{\nu_1}\dots x^2_{\nu_k},\ 
A^{(k)}=\!\!\!\!\!\sum_{\nu_1<\dots<\nu_k}\!\!\!\!\!x^2_{\nu_1}\dots x^2_{\nu_k}\;\label{co}.
\end{gather}
${M_\mu}$ are related to mass and NUT parameters, ${a_k}$ to angular momentum, and ${\lambda}$ is 
proportional to the cosmological constant.

The metric \eqref{metric} is an Einstein space obeying the equation
\begin{equation}\label{EE}
{R_{ab}=(D-1)\lambda g_{ab}}.
\end{equation}
It possesses
${m+1}={D-n}$ Killing vectors ${\partial_{k}}$, a $(D-2)$-rank Killing--Yano tensor \cite{KF},
and ${n}$ second-rank Killing tensors \cite{QuadrConst}, as well as
${n-2}$ higher-rank Killing tensors \cite{PKVK}.

The aim of this paper is to demonstrate that in the coordinates
${x_\mu,\psi_k}$,  both the Hamilton--Jacobi and Klein--Gordon
equations separate. To prove this we shall need a set of algebraic
relations which are valid for quantities which enter the metric
\eqref{metric}. It is useful to introduce quantities
\begin{equation}\label{Udef}
  U\equiv\prod_{\substack{\mu,\nu=1\\\mu<\nu}}^n(x_\mu^2-x_\nu^2)\;,\quad
  \upU_\mu\equiv\frac{U}{U_\mu}\;,
\end{equation}
which satisfy the important identities
\begin{subequations}\label{Uids}\allowdisplaybreaks
\begin{align}
&\sum_{\mu=1}^n x_\mu^{2(n-1)} \upU_\mu = (-1)^{n-1}U\;,\label{Un-1id}\\
&\sum_{\mu=1}^n x_\mu^{2k}\, \upU_\mu   = 0    
         \quad\text{for}\quad k=0,\dots,n-2\;,\label{Ukid}\\
&\sum_{\mu=1}^n \frac{1}{x_\mu^2}\, \upU_\mu  = \frac{U}{A^{(n)}}\;,\label{U-1id}\\
&\sum_{\mu=1}^n \frac{A_\mu^{(k)}}{x_\mu^2} \upU_\mu  = \frac{A^{(k)}}{A^{(n)}}U
         \quad\text{for}\quad k=0,\dots,n-1\;,\label{UA-1id}
\end{align}
\end{subequations}
and
\begin{equation}\label{upUder}
\partial_\mu\upU_\mu=0\;.
\end{equation}
The first two identities follow from the fact that the matrix 
${B^{\,\mu}_{(k)}=(-x_\mu^2)^{n\!-\!1\!-\!k}/U_\mu}$ is the inverse of ${A_\mu^{(k)}}$,
\begin{equation}\label{Ainverse}
\begin{gathered}
\sum_{k=0}^{n-1}\frac{(-x_\mu^2)^{n\!-\!1\!-\!k}}{U_\mu}A_\nu^{(k)}=\delta_\mu^\nu\;,\quad
\sum_{\mu=1}^{n} \frac{(-x_\mu^2)^{n\!-\!1\!-\!k}}{U_\mu}A_\mu^{(l)} = \delta_k^l\;
\end{gathered}
\end{equation}
(set ${l=0}$ in the last expression), \eqref{U-1id} follows from \eqref{Un-1id}
by substitution ${x_\mu\to1/x_\mu}$, and \eqref{UA-1id} can be verified using 
\eqref{U-1id}, \eqref{Ainverse}
and ${A_\mu^{(k)}=A^{(k)}-x_\mu^2 A_\mu^{(k-1)}}$. The identity \eqref{upUder} is obvious. 

The function ${U}$ is simply related to the determinant of the metric
\begin{equation}\label{det}
g={\det(g_{ab})}=\bigl(-c A^{(n)}\bigr)^\eps\, U^2\;.
\end{equation}

\section{Separability of the Hamilton--Jacobi equation}
The Hamilton--Jacobi equation for geodesic motion  
on a manifold with metric ${g_{ab}}$ has the form
\begin{equation}\label{HJ}
\frac{\partial S}{\partial\lambda}+g^{ab}\, \partial_a S\;\partial_b S=0\;.
\end{equation}
Here ${\lambda}$ denotes an `external' time which turns out to be 
an affine parameter of the corresponding geodesic motion. 
We want to demonstrate that in the background (\ref{metric}) 
the classical action $S$ allows a separation of variables 
\begin{equation}\label{sep}
S=-w\lambda + \sum_{\mu=1}^n S_{\mu}(x_{\mu})+ \sum_{k=0}^{m} \psic_k\psi_k
\end{equation}
with functions ${S_\mu(x_\mu)}$ of a single argument ${x_\mu}$.

Substituting \eqref{sep} into the Hamilton--Jacobi equation~\eqref{HJ} 
and multiplying by ${U}$ introduced in \eqref{Udef}, we obtain
\begin{equation}\label{HJsep}
\sum_{\mu=1}^{n}\upU_\mu F_{\mu} = w U +\eps\frac{\psic_n^2}{c}\frac{U}{A^{(n)}}\;,
\end{equation}
where $F_{\mu}$ is a function of $x_{\mu}$ only,
\begin{equation}\label{Fdef}
F_{\mu}=X_{\mu}S_{\mu}'^2+\frac{1}{X_{\mu}}
  \Bigl(\sum_{k=0}^m\bigl(-x_{\mu}^2\bigr)^{n\!-\!1\!-\!k}\,\psic_k\Bigr)^2\;.
\end{equation}
Here, the prime denotes the derivative of ${S_\mu}$ with respect to its single argument ${x_\mu}$.
Thanks to the identities \eqref{Uids}, the equation \eqref{HJsep} is satisfied if the
functions ${F_\mu}$ have the form
\begin{equation}\label{F=Sumconst}
F_{\mu}= \sum_{k=0}^{m} \xc_k\, (-x_\mu^2)^{n\!-\!1\!-\!k}\;,
\end{equation}
where ${\xc_k}$, ${k=1,\dots,n-1}$ are arbitrary constants, ${\xc_0=w}$, and the constant 
${\xc_n}$, which is present only in odd number of dimensions, is related to ${\psic_n}$ as
\begin{equation}\label{constn}
\xc_n=-\frac{\psic_n^2}{c}\;.
\end{equation}
The condition \eqref{F=Sumconst} leads to equations for ${S_\mu'}$
\begin{equation}\label{Scond}
  S_\mu'^2=-\frac1{X_\mu^2}\Bigl(\sum_{k=0}^m\bigl(-x_{\mu}^2\bigr)^{n\!-\!1\!-\!k}\psic_k\Bigr)^{\!2}\!
+\frac{1}{X_\mu}\sum_{k=0}^{m} \xc_k\, (-x_\mu^2)^{n\!-\!1\!-\!k}\!,
\end{equation}
which can be solved by quadratures. Notice that in odd dimensions there is
an additional term in which ${\xc_n}$ is not an independent constant, cf.\ Eq.~\eqref{constn}. 

Thus we have shown that Hamilton--Jacobi equation~\eqref{HJ} in the gravitational 
background (\ref{metric})
can be solved by the classical action ${S}$ in the separated form \eqref{sep} 
with ${S_\mu}$ satisfying \eqref{Scond}. The solution contains ${D}$ constants, namely
${\xc_0=w}$, ${\xc_1,\,\dots,\,\xc_{n-1}}$, and ${\psic_0,\,\dots,\,\psic_m}$.

The gradient of ${S}$ gives the momentum ${p_a=\partial_a S}$. Substituting 
our expression for ${S}$ we obtain ${p_a}$ in terms of the constants ${\xc_k}$ and ${\psic_k}$.
These relations can be inverted. 
Clearly, ${\psic_k=p_k}$ are constants linear in the momentum generated
by Killing vectors. 
To evaluate ${\xc_k}$ we rewrite \eqref{F=Sumconst} as
\begin{equation}\label{F-phin}
  \frac{F_\mu}{U_\mu}-\eps\frac{p_n^2}{cU_\mu x_\mu^2}
  =\sum_{k=0}^{n-1}\xc_k\frac{(-x_\mu^2)^{n\!-\!1\!-\!k}}{U_\mu}\;.
\end{equation}
It can be inverted using \eqref{Ainverse}. 
Employing the expression for ${\xc_n}$ with ${\psic_n=p_n}$ and the identity \eqref{UA-1id} we obtain 
\begin{equation}\label{const=SumF}
  \xc_k=\sum_{\mu=1}^n A_\mu^{(k)} \frac{F_\mu}{U_\mu}
     -\eps p_n^2\;\frac{A^{(k)}}{cA^{(n)}}\;,
\end{equation}
where ${F_\mu}$ is given by \eqref{Fdef} with ${p_\mu}$ and ${p_k}$ substituted for
${S_\mu'}$ and ${\psic_k}$, respectively.  

We thus found that the constants ${\xc_k}$ are quadratic in the momenta ${p_a}$
(for example, for ${k=0}$ we get ${w=\xc_0=g^{ab}p_ap_b}$).
It can be shown that they are the same as the constants introduced recently
using the Killing--Yano tensor 
and that they are generated by second rank Killing tensors \cite{QuadrConst}.

\section{Separability of the Klein--Gordon equation}
The behavior of a massive scalar field $\Phi$ in the gravitational background $g_{ab}$ is governed by
the Klein--Gordon equation
\begin{equation}\label{KG}
\Box\Phi=\frac{1}{\sqrt{|g|}}\,\partial_{a}(\sqrt{|g|}g^{ab}\partial_{b}\Phi)=m^2\Phi.
\end{equation}
This equation remains valid for the non-minimal coupling case as well. 
The term ${\xi R}$ is constant in the Einstein spaces 
and can be included into the definition of ${m^2}$.

Now, we demonstrate that the Klein--Gordon equation (\ref{KG}) in the background
(\ref{metric}) allows a multiplicative separation of variables
\begin{equation}\label{multsep}
\Phi=\prod_{\mu=1}^nR_{\mu}(x_{\mu})\prod_{k=0}^{m}e^{i\psic_k\psi_k}.
\end{equation}

This equation has the following explicit form
\begin{eqnarray}\label{KGm}
\sqrt{|g|}m^2\Phi&=&\sum_{\mu=1}^n\partial_{\mu}\Bigl(\frac{\sqrt{|g|}}{U_{\mu}}X_{\mu}\partial_{\mu}\Phi\Bigr)\nonumber\\
&+& \sum_{\mu=1}^n\frac{\sqrt{|g|}}{U_{\mu}X_{\mu}}\Bigl(\sum_{k=1}^{m}(-x_{\mu}^2)^{n-1-k}\partial_k\Bigr)^{\!2}\Phi\nonumber\\
&-&\eps\frac{\sqrt{|g|}}{cA^{(n)}}\partial^2_n\Phi\;.
\end{eqnarray} 
Here we used the quasidiagonal property of the inverse metric $g^{ab}$ and the fact 
that $\partial_k$ are Killing vectors. We further notice that
\begin{equation}
\sqrt{|g|}\propto UP^{\eps},\quad P\equiv \prod_{\mu=1}^n x_{\mu},
\end{equation}
where ``$\propto$'' means equality up to a constant factor (which can be ignored in Eq.~\eqref{KGm}).
Using the identities \eqref{Un-1id}, \eqref{U-1id}, (\ref{upUder}) and the definition 
of $\upU_{\mu}$ we find that (\ref{KGm}) gives
\begin{align}\label{22}
&\sum_{\mu=1}^n\upU_{\mu}\left[(-1)^nm^2x_{\mu}^{2(n-1)}\Phi+
\partial_{\mu}(P^{\eps}X_{\mu}\partial_{\mu}\Phi)/P^{\eps}\right]\;\\
&+\sum_{\mu=1}^n\biggl[\frac{\upU_{\mu}}{X_{\mu}}\Bigl(\sum_{k=1}^{m}(-x_{\mu}^2)^{n-1-k}\partial_k\Bigr)^{\!2}\Phi-
\frac{\eps\upU_{\mu}}{cx_{\mu}^2}\partial_{n}^2\Phi\biggr]=0\;.\nonumber
\end{align} 
Using the ansatz (\ref{multsep}) we find
\begin{equation}
\partial_k\Phi=i\psic_k\Phi,\quad \partial_{\mu}\Phi=\frac{R_{\mu}'}{R_{\mu}}\Phi,
\quad  \partial^2_{\mu}\,\Phi=\frac{R_{\mu}''}{R_{\mu}}\,\Phi,
\end{equation}
and the Klein--Gordon equation (\ref{22}) takes the form
\begin{equation}\label{KGsep}
\sum_{\mu=1}^n\upU_{\mu}G_{\mu}\Phi=0,
\end{equation}
where $G_{\mu}$ is function of $x_{\mu}$ only,
\begin{eqnarray}
G_{\mu}&=&(-1)^nm^2x_{\mu}^{2(n-1)}+\frac{R'_{\mu}}{R_{\mu}}\Bigl(X_{\mu}'+\eps\frac{X_{\mu}}{x_{\mu}}\Bigr)
+X_{\mu}\frac{R_{\mu}''}{R_{\mu}}\nonumber\\
&-&\frac{1}{X_{\mu}}\Bigl(\sum_{k=1}^{m}(-x_{\mu}^2)^{n-1-k}\psic_k\Bigr)^{\!2}+\frac{\eps\psic_n^2}{cx_{\mu}^2}.
\end{eqnarray} 
As earlier, the prime means the derivative of functions 
${R_\mu}$ and ${X_\mu}$ with respect to their single argument ${x_\mu}$.
Employing the identity \eqref{Ukid} we realize that (\ref{KGsep}) is automatically satisfied when
\begin{equation}
G_{\mu}=\sum_{k=1}^{n-1} b_{k}(-x_\mu^2)^{n\!-\!1\!-\!k}\;,
\end{equation}
where $b_k$ are arbitrary constants.

Therefore we have demonstrated that the Klein--Gordon equation (\ref{KG}) in the 
background (\ref{metric}) allows a multiplicative separation
of variables (\ref{multsep}), where functions $R_{\mu}(x_{\mu})$ satisfy the ordinary 
second order differential equations
\begin{align}\label{ODE}
&\bigl(X_{\mu}R_{\mu}'\bigr)'+\eps\frac{X_{\mu}}{x_{\mu}}R_{\mu}'-
\frac{R_{\mu}}{X_{\mu}}\Bigl(\sum_{k=0}^m(-x_{\mu}^2)^{n-1-k}\psic_k\Bigr)^{\!2}\nonumber\\
&\mspace{120mu}-\sum_{k=0}^m b_{k}(-x_{\mu}^2)^{n\!-\!1\!-\!k}R_{\mu}=0\;.
\end{align}
Here $b_0=m^2$,  $b_1\dots b_{n-1}$ are arbitrary separation constants. 
The constant $b_n$ is present only in an odd number of 
spacetime dimensions and is related to the constants $\psic_n$ and $c$ through
\begin{equation}
b_n=\frac{\psic_n^2}{c}.
\end{equation}
We expect that the separation constants are related to the Killing tensors obtained in \cite{QuadrConst}.


\section{Discussion}
We demonstrated the separability of the Hamilton--Jacobi and the scalar field
equations in the general (higher-dimensional) Kerr--NUT--AdS spacetime. 
For particle motion the separability implies that the corresponding
equations of motion can be written in the first order form \eqref{Scond}.
In the Klein--Gordon case we obtained a set of ordinary second order 
differential equations (\ref{ODE}). The problem to solve them is usually much simpler.
Even when some of these
equations cannot be solved in terms of known elementary or special
functions, one can always use numerical methods. The
numerical integration of ordinary
differential equations can be performed very effectively. 

In the present paper we established the separability property by 
`brute force'---by writing the corresponding equations in a special
coordinate system.  As we already mentioned, the constants of
separation are directly related to the existing complete set of the
second rank Killing tensors \cite{QuadrConst}. It would be interesting
to derive the separability property by starting with the general
symmetry properties of the considered spacetime, using, for example,
the results of \cite{BeFr:79}.

The separation of variables in the scalar field equation can be used
for the study of different interesting problems. One of them is the
calculation of the bulk Hawking radiation of higher dimensional
rotating black holes. As it was shown by Teukolsky \cite{Teuk_a, Teuk_b}
in the 4D Kerr metric, not only the scalar field equation allows
separation of variables, but the equations of the other (massless)
fields with non vanishing spin can also be decoupled and separated.
An interesting question is whether the existing symmetry connected
with a complete set of the Killing tensors in the  general
Kerr--NUT--AdS spacetime makes such a decoupling and
separation of the higher spin fields equations possible.
 
\vspace*{-3ex}
\section*{Acknowledgments}
\vspace*{-1ex}

V.F. thanks the Natural Sciences and Engineering
Research Council of Canada and the Killam Trust for the financial
support. P.K. was kindly supported by the grant GA\v{C}R 202/06/0041 
and appreciates the hospitality of the University of Alberta.
D.K. is grateful to the Golden Bell Jar Graduate
Scholarship in Physics at the University of Alberta.
The authors also thank Don~N.~Page and M.~Vasudevan for reading the manuscript. 
\appendix


\end{document}